# Quasi-Integrable Nonlinear Optics Experiment at IOTA

S. A. Antipov[1, a)], S. Nagaitsev[2], and A. Valishev[2]

[1]The University of Chicago, 5720 S. Ellis Ave, Chicago, IL 60637
[2]Fermilab, Batavia, IL 60510

a)Corresponding author: antipov@uchicago.edu

**Abstract.** At Integrable Optics Test Accelerator it is possible to create a nonlinear focusing optics with one invariant of motion using just conventional magnets. 6D simulations show that this will allow to achieve a tune spread of 0.05 without significant reduction of dynamic aperture.

## INTRODUCTION

Contemporary accelerators are designed with linear focusing lattices. In ideal case, in such optics single particle motion is fully integrable and thus unconditionally stable. In reality, machines always have nonlinearities, resulting from magnet imperfections or specially introduced (e.g., sextupoles for chromaticity correction). These nonlinearities break integrability of the Hamiltonian, resulting in resonant behavior and particle losses.

In 2010 Danilov and Nagaitsev discovered a nonlinear integrable focusing lattice [1]. It is predicted to create a spread of betatron frequencies of the order of unity [2], which provides significant suppression of collective instabilities via Landau damping. Recent theoretical studies [3] also show the suppression of space charge driven parametric resonances and associated beam loss via halo formation.

Fermilab is building an Integrable Optics Test Accelerator (IOTA), a part of Advanced Superconducting Test Accelerator (ASTA) facility, to test the concept of nonlinear integrable focusing lattice in realistic setup of a storage ring.

## INTERGABLE OPTICS TEST ACCELERATOR

IOTA is a storage ring, designed to operate with 50-150 MeV electrons or protons of the same momentum. It consists of two 1.8m-long sections with equal beta-functions for nonlinear magnets, and so called T-inserts with transfer matrices of thin axially symmetric lenses between these sections (Fig. 1). The main parameters of the ring are summarized in Table 1.

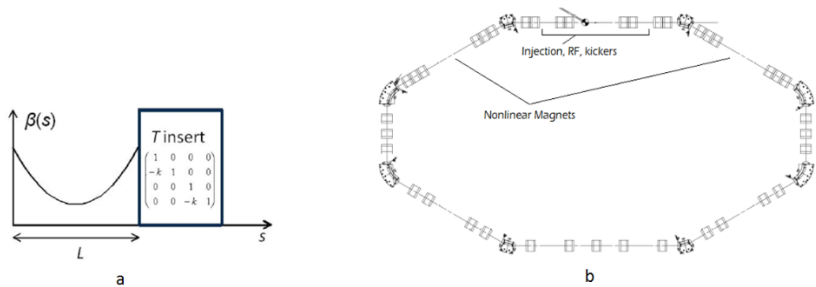

**FIGURE 1.** Element of periodicity of IOTA– (a) and a drawing of the ring – (b)

The nonlinear magnets create a potential, which can be described as:

$$U(x, y) = t \cdot \text{Re}\left[(x+iy)^2 + \frac{2}{3c^2}(x+iy)^4 + \frac{8}{15c^4}(x+iy)^6 + \frac{16}{35c^6}(x+iy)^8 + ...\right], \quad (1)$$

where $x$ and $y$ denote normalized coordinates, $t$ and $c$ are parameters. For IOTA nonlinear magnets $c^2 = 1$ cm.

**TABLE 1.** IOTA parameters

| Property | Value |
|---|---|
| Electron energy | 150 MeV |
| Ring circumference | 40 m |
| Length of NL magnets | 1.8 m |
| Phase advance per NL section | 0.3 |
| Synchrotron damping time | ~1 sec |
| Equilibrium RMS beam size | 0.1 mm |
| Chromaticity: x, y | -10, -7 |
| RF harmonic number | 4 |
| Bunch length | 5 cm |
| Tunes: x, y, s | 5.3, 5.3, 3·10$^{-3}$ |

Figure 2 depicts magnetic field lines inside the magnet. Because it has two points of singularity at $x = \pm c$, design, manufacturing, and testing of the nonlinear magnets is challenging [4]. A possible solution is to approximate the nonlinear potential (1) using a set of conventional magnets. For example, if one can take the octupole term in (1), which is the lowest order term that creates a nonlinear tune spread. To create this potential one needs to scale octupole strength as $\beta(s)^{-3}$. Then the resulting Hamiltonian:

$$H = \frac{1}{2}(x^2 + y^2 + p_x^2 + p_y^2) + \frac{2t}{3c^2}(x^4 - 6x^2y^2 + y^4) \quad (2)$$

is independent of longitude coordinate $s$ and thus it is an invariant of motion. Experiment with octupoles is the first one to be conducted at IOTA.

The Hamiltonian (2) takes into account betatron oscillations only. Previous works on simulation ([2], [5]) of dynamics in IOTA also did not consider synchrotron oscillations, which can affect the invariants of motion. Chromaticities of phase advance and beta-functions break the assumptions of [1], changing both the transfer matrix and beta-functions in nonlinear inserts in Fig. 1 (a). To address this problem a 6D simulations was required.

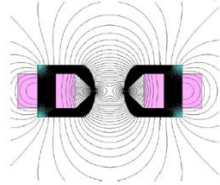

**FIGURE 2.** Magnetic field lines inide IOTA nonlinear magnets [5]

## 6D SIMULATIONS

Our simulation setup included dipole magnets, quadrupoles, and RF cavity. The model featured both betatron and synchrotron oscillations and had the chromaticity of the real machine. A single nonlinear insert with octupole potential was created by 18 10 cm long magnets.

Simulation were performed in Lifetrack code, which combines tools for 6D tracking and Frequency Map Analysis. Figure 3 depicts FMA footprints of the phase space, with the amplitudes of oscillations being measured at the center of the nonlinear section. For amplitudes lower 2.5 mm no crossing of resonant lines (red) is observed, and the motion is bounded. The corresponding tune spread is 0.03.

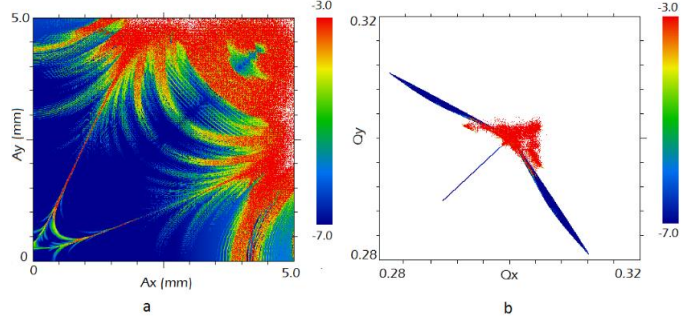

**FIGURE 3.** Amplitude – (a) and frequency – (b) phase space footprints for a ring with 1 octupole section. In color is the log of relative deviation of betatron frequency during FMA sampling window. $t = 0.4$, $c^2 = 1$ cm.

A time frame of 1 sec – characteristic time of synchrotron damping – has been simulated with 6D tracking. No particle loss detected for amplitudes lower 2.5 mm. Relative deviation in of the Hamiltonian (2) is less or of the order of $10^{-3}$ (Fig. 4). These small oscillations are attributed to the synchrotron motion, but overall the Hamiltonian is conserved. It should be noted that in IOTA the chromaticities in x- and y-planes almost match (see Table 1), which is the best case for single-particle stability according to [6]; effect of significant mismatch in chromaticities on particle dynamics is yet to be studied.

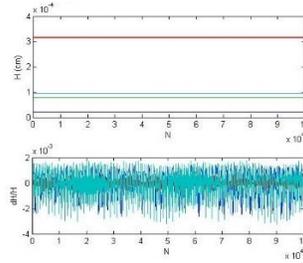

**FIGURE 4.** Hamiltonian and its relative deviation for different particles.

It is intuitive that the tune spread depends on the strength of octupoles – parameter $t$ in (1). As t increases particles oscillating with bigger amplitudes get higher frequency shift, but the dynamic aperture decreases also. As a result, maximum achievable tune spread reaches its maximum at $t = 0.4$-$0.5$ (Fig. 5). Further increase of octupole strength leads only to reduction of dynamic aperture. In all subsequent simulations $t = 0.4$ was used.

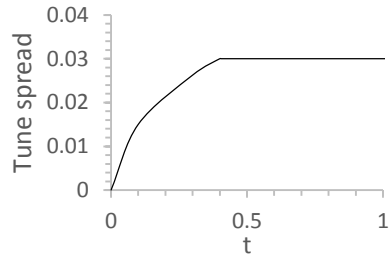

**FIGURE 5.** Dependence of maximum tune spread on strength parameter $t$ for 1 octupole insert. $c^2 = 1$ cm

In order to create a larger tune spread a large one can include a quadrupole term of (1). Then x- and y-tunes become different, and the machine no longer operates in a difference resonance. These way one can achieve a tune spread of up to 0.05 (Fig. 6). The downside of this approach is the reduction of dynamic aperture in one planes.

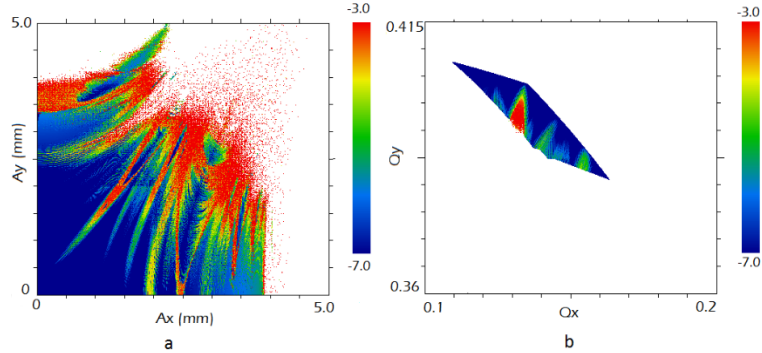

**FIGURE 6.** Amplitude – (a) and frequency – (b) phase space footprints for a ring with 1 quadrupole+octupole section; t = 0.4, c² = 1 cm.

# FUTURE EXPERIMENT

During the first stage, IOTA will operate with 150 MeV pencil electron beam to test the dynamics in the ring. Shortly after injection to the ring from ASTA linac, the transverse size of electron bunch will shrink to its equilibrium value of ~ 0.1 mm due to synchrotron radiation. This transverse size is the major source of error in the experiment, limiting the precision of measurement of betatron amplitude.

When the transverse oscillations decayed one can use two stripline kickers: horizontal and vertical, which operate in a wide range of kick voltages, to create an arbitrary initial displacement in the phase space. After the initial kick evolution of particle trajectories can be studied using Poincare mapping technique. To create Poincare maps one can measure position of particles with 20 beam position monitors, distributed throughput the ring. The BPMs have 0.1 mm turn-by-turn and 1 μm closed orbit resolution.

Measuring particles' positions of each turn and performing Fourier transform, one can obtain dependence of betatron tune on the amplitude of oscillations, as well as maximum tune spread and dynamic aperture. Whereas previous similar experiments of nonlinear dynamics, such as [7], used only natural nonlinearities already present in the machine, we will be able to control the strength of the nonlinearity in the ring.

Sources of error in the experiment include imperfections of both linear optics of the ring and octupole potential. By design, IOTA provides a huge accuracy of beta-functions and phase advances: $\delta\beta/\beta < 10^{-2}$, $\delta\psi < 10^{-3}$, and it is reasonable to assume that relative error in octupole potential does not exceed 0.1. For these ranges of errors, our simulations show no significant reduction of dynamic aperture or achievable tune spread. Overall, the absolute uncertainties in the experiment will be of the order of 0.1 mm in betatron amplitudes and $10^{-3}$ in tunes.

# CONCLUSION

Is this paper we have described the first planned experiment with quasi-integrable nonlinear optics at IOTA. Using conventional magnets such as octupoles one can achieve a preservation of one integral of motion in a highly nonlinear focusing lattice. 6D simulations of the actual machine have shown that synchrotron oscillations and chromaticity do not destroy particle motion. Our results predict an achievable tune spread of 0.05 without significant reduction of dynamic aperture.